\documentclass{desyproc}

\begin{document}
\title{Reflective Elastic Scattering at LHC}

\author{{\slshape Sergey Troshin, Nikolay Tyurin}\\[1ex]
IHEP, Protvino, 142281 Russia}

\contribID{smith\_joe}

\desyproc{DESY-PROC-2009-xx}
\acronym{EDS'09} 
\doi  

\maketitle

\begin{abstract}
 We discuss
effects of reflective scattering for hadron and heavy nuclei
collisions at the LHC and asymptotical energies. It is shown that
the reflective scattering might lead to decreasing matter density
with energy beyond the LHC energies. Limiting form of energy
dependence of hadron density is obtained. Unitarity upper bound for the absolute value
of the real part of elastic scattering amplitude and two-particle inelastic binary reactions amplitudes
in impact parameter representation is two times less than the corresponding
 bound for the imaginary part of the elastic scattering amplitude.
The former limit restricts a possible odderon contribution.
\end{abstract}

\section{Reflective scattering}

A new physical interpretation of unitarity saturation in elastic
scattering  as a reflective scattering was proposed in \cite{mirror} proceeding from optical analogy.
This interpretation is
related to the non-perturbative aspects of strong interactions and
follows from the  specific property of the unitarity saturation
when elastic $S$-matrix becomes negative and $S(s,b)|_{b=0}\to -1$ at $s\to \infty$. It
should be noted that $S(s,b)=1+2if(s,b)$, where $f(s,b)$ is the
elastic scattering amplitude in the impact parameter
representation.

In particular, we would like to note that the possible values of elastic $S$ matrix can be negative
(in the pure imaginary case).
Transition to the reflective scattering mode is naturally
 reproduced by the $U$-matrix form of elastic unitarization.
The elastic scattering $S$-matrix ($2\to 2$ scattering matrix element)
in the impact parameter representation is
written in this unitarization scheme in the form of linear fractional transform:
\begin{equation}
S(s,b)=\frac{1+iU(s,b)}{1-iU(s,b)}, \label{um}
\end{equation}
where $U(s,b)$ is the generalized reaction matrix, which is
considered to be an input dynamical quantity.
 For simplicity we consider
 the case of pure imaginary $U$-matrix and make the replacement $U\to iU$ in (\ref{um}).
The reflective scattering mode ($S(s,b)<0$) starts to appear at
the energy $s_R$, which is determined as a solution of the equation
$U(s_R,b=0)=1$.
At $s>s_R$ the elastic scattering  acquires ability for
reflection, while  inelastic overlap function $h_{inel}(s,b)$ gets a
 peripheral impact parameter dependence in the region $s>s_R$.
 It should be noted that unitarity condition for the elastic scattering amplitude $F(s,t)$,
 which can be written in the form
 \begin{equation}\label{un}
 \mbox{Im}F(s,t)=H_{el} (s,t)+H_{inel} (s,t),
 \end{equation}
 where $H_{el,inel}(s,t)$ are the corresponding elastic  and inelastic overlap function
 introduced by Van Hove \cite{vanh}.
The functions $H_{el,inel}(s,t)$ are related to the functions
$h_{el,inel}(s,b)$ via the Fourier-Bessel transforms, i.e.
\begin{equation}\label{hel}
H_{el,inel} (s,t)=\frac{s}{\pi^2}\int_{0}^{\infty} bdb h_{el,inel}(s,b) J_0(b\sqrt{-t}).
\end{equation}
The elastic and inelastic cross--sections can be obtained as
follows:
\begin{equation}\label{selin}
\sigma_{el,inel}(s)\sim \frac{1}{s} H_{el,inel} (s,t=0).
\end{equation}
Saturation of unitarity leads to the
 peripheral dependence of $h_{inel}(s,b)$. It is a manifestation
 of the self--damping of the inelastic channels
at small impact parameters.
The function $h_{inel}(s,b)$ reaches its maximum
 value at $b=R(s)$, note that \[R(s)\sim \frac{1}{M}\ln s,\]
 while an elastic scattering (due to reflection) occurs effectively at smaller values
of impact parameter, i.e.
$\langle b^2 \rangle_{el}<\langle b^2 \rangle_{inel}$.
At the  values of energy $s>s_R$ the equation $U(s,b)=1$ has a solution in the
physical region of impact parameter values, i.e. $S(s,b)=0$ at $b=R(s)$.
Fig. 1 shows
the regions where elastic $S$-matrix has positive and
negative values.
\begin{figure}[hbt]
\begin{center}
\includegraphics[scale=0.5]{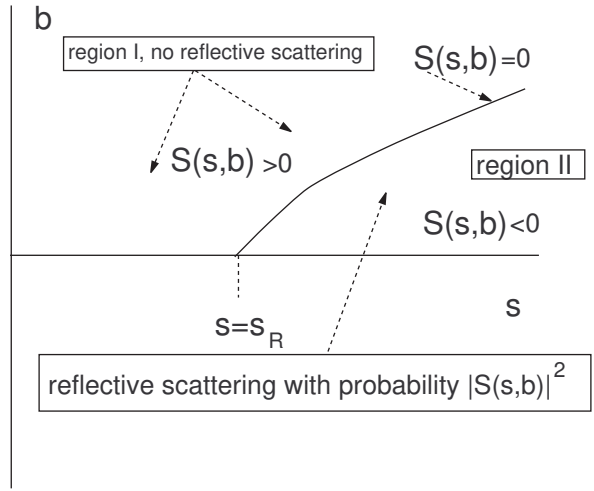}
\caption{\small\it Regions of positive (absorptive scattering) and
negative values (absorptive and reflective scattering) of the
function $S(s,b)$ in the $s$ and $b$ plane.}
\end{center}
\end{figure}
Of course, the reflective scattering exists only in the elastic channel. For example, all inelastic
binary reactions have amplitudes $\tilde f(s,b)$ in the impact parameter representation, which satisfy to
inequality
$|\tilde f(s,b)|\leq 1/2$ while elastic scattering amplitude satisfies to $|f(s,b)|\leq 1$. It follows
from unitarity equation in the impact parameter representation:
\[
\mbox{Im} f(s,b)=h_{el}(s,b)+h_{inel}(s,b);
\]
amplitudes $\tilde f(s,b)$ contribute to $h_{inel}(s,b)$.
 Upper bound for
the elastic scattering cross--section is four times higher than the upper bound for the inelastic cross--section
as it was recently  demonstrated in \cite{Martin}. Scattering dynamics in the elastic channel such as $pp\to pp$
is therefore strikingly different, e.g. from the inelastic binary diffractive process, such as $pp\to pN$,
where $N$ is an isobar. Latter reactions should have a peripheral impact parameter profile, which can be related
to the dominating contribution of helicity-flip amplitudes. The unitarity limit and black disk
limit are the same for the inelastic overlap function, but those limits are different
for the elastic overlap function
\[ h_{el}(s,b)\equiv |f(s,b)|^2. \] Unitarity limit for elastic overlap
function is four times higher than the black disk limit. This is an important point for consideration of
exclusive limit of inclusive reactions.
Saturation of unitarity  leads to suppression of inelastic
cross-section, i.e. at fixed impact parameter ($b<R(s)$) $h_{inel}(s,b)\to 0$ at $s\to\infty$ and
\begin{equation} \label{tasel}
\sigma_{el}(s)\sim R^2(s),\;\;\; \sigma_{inel}(s)\sim R(s).
\end{equation}
Thus, $H_{el}(s,t)$ which has the following $t$--dependence
 \[ H_{el}(s,t) \sim \frac {RJ_1(R\sqrt{-t})}{\sqrt{-t}},\]
dominates over $H_{inel}(s,t)$, which depends on $t$ like
 \[
 H_{inel}(s,t) \sim RJ_0(R\sqrt{-t}),
\]
  at $-t=0$, but it is not the
case for the scattering in the non-forward directions. In this region these
two functions have similar energy dependencies proportional to $R^{1/2}(s)$ at
rather large fixed values of $-t$.
The mean  impact parameter values for elastic and inelastic interactions have also similar
energy dependencies
\begin{equation} \label{taseln}
{\langle b^2\rangle}_{el}(s)\sim R^2(s),\;\;\; {\langle
b^2\rangle}_{inel}(s)\sim R^2(s),
\end{equation}
but the   value of impact parameter averaged over all interactions
\[
{\langle b^2\rangle}_{tot}(s)=\frac{\sigma_{el}(s)}{\sigma_{tot}(s)}{\langle b^2\rangle}_{el}(s)+
\frac{\sigma_{inel}(s)}{\sigma_{tot}(s)}{\langle b^2\rangle}_{inel}(s)
\]
acquires the main contribution  from elastic scattering according to Eq. {\ref{tasel}}.
Therefore, the inelastic intermediate states will give subleading  contribution
to the slope of diffraction cone $B(s)$,
\[
B(s)\equiv \frac{d}{dt}\ln (\frac {d\sigma}{dt})|_{t=0},
\]
at asymptotical energies.
Indeed, since
$B(s) \sim {\langle b^2\rangle}_{tot}(s)$,
it can be written in the form
\[
B(s)=B_{el}(s)+B_{inel}(s),
\]
where
$B_{el}(s)\sim R^2(s)$,
 while $B_{inel}(s)
 \sim R(s)$. It should be noted that
  both terms $B_{el}(s)$ and $B_{inel}(s)$ are
 proportional to $R^2(s)$ in case of the absorptive scattering.

Under reflective scattering, behavior of the function
$H_{inel}(s,t)$ is determined by a peripheral impact parameter
profile and its $-t$ dependence is different. Meanwhile,
 the elastic overlap function $H_{el}(s,t)$
  has similarities with that function in
the case of approach where absorption is only presented.  As a result, zeroes
and maxima of the functions $H_{el}(s,t)$ and $H_{inel}(s,t)$
 will be located at different
values of $-t$ and zeroes and maxima of $\mbox{Im}F(s,t)$ will  also  be located
at different position in the cases of absorptive and the reflective scattering.
In the case of reflective scattering, dips and maxima will be  located in the region of
lower values of $-t$.
 We would like to note that the presence of reflective scattering enhances
 the large $-t$ region by factor
  $\sqrt{-t}$ compared to absorptive scattering.
 Despite that these two mechanisms lead at the asymptotics to
 the significant differences in the total, elastic and
 inelastic cross-section dependencies, their predictions for the differential cross-section
 of elastic
  scattering are  not so much different at small and moderate values of $-t$.
\section{Intermittent remark on unitarity and real part of scattering amplitude}

It is evident that there are serious  difficulties in accounting
all  known
 dynamical issues and limitations into a particular phenomenological model. But it  is equally
 difficult to expect
 that the model inconsistent with unitarity (i.e. the one violating probability
 conservation law) would adequately  reflect  the  dynamics of hadron interaction and
 provide reliable predictions.
 To fulfill unitarity condition under a model construction of the elastic
 amplitudes, it is natural to use unitarization approaches such as eikonal or $U$-matrix, which consider
  amplitudes in the  impact parameter space. They automatically guarantee that elastic amplitude in the impact
  parameter representation will obey unitarity condition.

 Despite that the full implementation of unitarity is not possible nowadays  (cf. e.g. \cite{land}), the
 amplitude in the impact parameter space should not exceed unity anyway.
 However,  when the amplitude $F(s,t)$ is constructed in
  the $s$ and $t$ representation, it is a priori not evident that the particular form
  of this amplitude being
   transformed into the impact parameter space $f(s,b)$  would satisfy unitarity.
  This remains to be true, even when the
  model under consideration leads to the predictions for observables
   which  explicitly agree with axiomatic bounds, e.g.
 such as  well known Froissart-Martin bound for the total
 cross--sections. Getting agreement with experimental data at finite energies
 and asymptotical bounds at $s\to\infty$ is not enough since a wide class of
 functional dependencies can successfully describe experimental data
 and provide correct asymptotical behavior.
Additional steps to justify  that the impact parameter
 amplitude is {\it at least} less  than unity  in the whole region of kinematical variables are
 necessary.

 In the above remarks we supposed that imaginary part of scattering amplitude is a dominating one.
 Further unitarity restriction exists for models which do not suppose domination of imaginary
 part of scattering amplitude, such as models with maximal odderon contribution \cite{mart}.

 Indeed, unitarity condition in the impact parameter representation for the elastic scattering amplitude can
 be rewritten in the form:
 \[
 \mbox{Im} f(s,b)[1-\mbox{Im} f(s,b)]=[\mbox{Re} f(s,b)]^2+h_{inel}(s,b).
 \]
Since $0\leq\mbox{Im}(s,b)\leq 1$, we obtain that unitarity  limits the real part
of scattering amplitude in the following
 way
 \[
[\mbox{Re} f(s,b)]^2\leq 1/4,
\]
\[
-\frac{1}{2}\sqrt{1-4h_{inel}(s,b)}\leq \mbox{Re} f(s,b)\leq \frac{1}{2}\sqrt{1-4h_{inel}(s,b)}.
\]
The function $\mbox{Re} f(s,b)$ can be sign changing one in contrast with $\mbox{Im}(s,b)$).
This limitation, as it was already mentioned, is essential for the models with odderon and is
indirectly in favor of the standard procedure of neglecting the real part of scattering amplitude compared
to its imaginary part. It  also is  evident that absolute value of the
real part and imaginary part of elastic scattering amplitude $f(s,b)$
cannot reach their maximal values simultaneously, moreover
when $ \mbox{Im} f(s,b)\to 1$, saturating unitarity limit at large values of s in
the region $b<R(s)$, then $\mbox{Re} f(s,b)\to 0$ in this kinematical region.
It should be noted that this saturation does not suppose
 that $\mbox{Re} f(s,b)$ vanish everywhere. It means
that $[\mbox{Re} f(s,b)]^2$ should have  a peripheral impact parameter profile.
 The same conclusion is valid when $ \mbox{Im} f(s,b)\to 1/2$,
saturating the black disk  limit at large values of s in
the region $b<R(s)$, then $\mbox{Re} f(s,b)\to 0$ because $h_{inel}(s,b)\to 1/4$ in this region.
The above difference
in the impact parameter profiles would result in the different energy dependencies
of $ \mbox{Im} F(s,t=0)$ and $ \mbox{Re} F(s,t=0)$ bringing maximal odderon on the edge
of contradiction with unitarity (or black disk) limit saturation.
Of course, unitarity or black disk limits saturation itself does
not follow
 from axiomatic field theory, but
 we would like to note,
that it is much more natural to expect that  it could be  a manifestation of
a maximal strength of strong interaction instead of behavior of the real
part of the forward scattering amplitude in the form $\mbox{Re}F(s,t=0)\sim s\ln^2 s$
as it happens in the models incorporating the maximal odderon regime.

\section{Reflective scattering and deconfinement}
Possible existence  of the reflective scattering at very high energies implies that
confinement becomes stronger and stronger as the collision energy increases and proton collisions
resemble more and more collisions of hard spheres. In this section  we address
   one aspect of the broad  problem of
transition to the deconfined state of matter, namely, we  discuss
the role of the reflective scattering on the energy dependence of
density in the percolation mechanism of the transition to the
deconfined state of matter.

The main idea of the percolation mechanism of deconfinement is a
formation in the certain volume of a connected hadron cluster due
to increasing temperature and/or hadron density \cite{satz}, i.e.
when vacuum as a connected medium disappears,  the deconfinement
takes place. This process has typical critical dependence on
particle density. Thus, it was proposed to use percolation to
define the states of matter and consider the disappearance of a
large-scale vacuum as the end of hadronic matter
existence\cite{satz,satz2}.

The probability of reflective scattering at $b<R(s)$ and $s> s_R$ is determined by the magnitude
 of $|S(s,b)|^2$; this probability is equal to zero at $s\leq s_R$ and $b\geq R(s)$ (region I on Fig.1).
At the energies $s> s_R$  reflective scattering will mimic presence of repulsive core in
 hadron and meson interactions.
  Presence of the reflective scattering can be
 accounted for  using van der Waals method (cf. \cite{cleym}).
 This approach was used originally for description of the  fluids behavior starting from
 the gas approximation by means of
  taking into account the nonzero size of molecules. Consider central collision of two identical nuclei having $N$ hadrons in total with center of mass energy
$\sqrt{s}$ per nucleon and calculate
hadron density $n_R(T,\mu)=N/V$ in the initial state at given
 temperature $T$ and baryochemical potential $\mu$ in the presence of
reflective scattering.
The effect of the reflective scattering of
  hadrons is equivalent  to decrease of the volume of the available  space
  which the hadrons are able to occupy in the case when reflective scattering is absent.
  Thus followings to van der Waals method, we must then replace volume $V$ by \[V-p_R(s)V_R(s)\frac{N}{2},\]
  i.e. we should write
  \[
  n(T,\mu)=\frac{N}{V-p_R(s)V_R(s)\frac{N}{2}},
\]
where $n(T,\mu)$ is hadron density  without account for reflective scattering and
 $p_R(s)$ is the averaged over volume $V_R(s)$ probability of reflective scattering:
\[p_R(s)=\frac{1}{V_R(s)}\int_{V_R(s)}|S(s,r)|^2 d^3x.\]
The volume $V_R(s)$ is determined by the radius of the reflective scattering.
Here we  assume spherical symmetry of hadron interactions,
i.e. we replace impact parameter $b$ by $r$ and approximate the volume $V_R(s)$ by
$V_R(s)\simeq (4\pi/3)R^3(s)$. Hence, the density $n_R(T,\mu)$  is connected
with corresponding density in the approach without reflective scattering $n(T,\mu)$
 by the following relation
\[
  n_R(T,\mu)=\frac{n(T,\mu)}{1+\alpha(s)n(T,\mu)},
\]
where $\alpha(s)={p_R(s)V_R(s)}/{2}$. Let us now estimate change of the function
$n_R(T,\mu)$ due to the presence of reflective scattering. We can
approximate $p_R(s)$ by the value of $|S(s,b=0)|^2$ which tends to
unity at $s\to\infty$. It should be noted that the value
$\sqrt{s_R}\simeq 2$ $TeV$ \cite{sr}. Below this energy there is
no reflective scattering, $\alpha(s)=0$ at $s\leq s_R$, and
therefore corrections to the hadron density are absent. Those
corrections are small when the energy is not too much higher than
$s_R$. At $s\geq s_R$ the value of $\alpha(s)$ is positive,
and presence of reflective
scattering diminishes hadron density. We should expect that this
effect would already be  noticeable at the LHC energy $\sqrt{s}\simeq 5$
TeV in $Pb+Pb$ collisions. At very high energies ($s\to\infty$)
\[
n_R(T,\mu)\sim 1/\alpha(s)\sim M^3/\ln ^3 s.
\]
This  limiting dependence for the hadron density  appears due to
 the presence of the reflective scattering which results in similarity of head-on hadron collisions
with  scattering of hard spheres. It
can be associated with  saturation of the Froissart-Martin bound for the total cross-section. It should be noted
that this dependence has been obtained under assumption on spherical symmetry of hadron interaction region.
Without this assumption, limiting dependence of the hadron density in transverse plane can only be obtained,
i.e. transverse plane density of hadrons would have then the following behavior
\[
n_R(T,\mu)\sim M^2/\ln ^2 s.
\]
To conclude this section, we would like to note that the lower densities of hadron matter are needed
for percolation (and transition to the deconfined state) in the presence of reflective scattering.
It might be useful to note that the rescattering processes
  are also affected by the reflective scattering.
Reflective scattering would lead to noticeable effects at the LHC energies and beyond and
could help in searches of the
deconfined state and studies of properties of transition mechanism to this state of matter which might
proceed by means of percolation. Thus, it will affect description of
initial state dynamics in nuclear
interactions at the LHC energies
 by introducing notion of limiting density of strongly interacting matter at respective energies.

\section*{Conclusion}
Thus, at very high energies  there would be two different  regions of
 impact parameter distances in particles collisions, namely the outer region
(peripheral collisions) where elastic scattering has exclusively a shadow origin and
inner region (central collisions)
where reflective and absorptive scattering give competing contributions, reflective scatteing
contribution increases while absorptive scattering contribution decreases at fixed impact parameter.
It is not surprising that the model with reflective scattering contribution leads to significantly
higher values for total and elastic cross--sections at the LHC
energies\footnote{The value of total cross--section  about 150 $mb$ is predicted for $pp$-interactions
 at the LHC starting up energy
$\sqrt{s}=7$ $TeV$.} while renders to
the standard values for the inelastic cross--section.
In the geometric terms, the generic scattering
 picture at fixed energy beyond the black disc limit can be described
 as a scattering off
a partially reflective and partially absorptive disk
surrounded by the black ring (which becomes grey at larger values of the
impact parameter). The evolution with energy  is characterized
by increasing albedo due to the  interrelated  increase of reflection
  and decrease of absorption at small impact parameters. This picture predicts
  that the scattering amplitude at
  the LHC energies is beyond the black disk limit at small impact
  parameters and it  provides  explanation for the regularities
  observed in cosmic rays studies, e.g. the existence of the knee
  in the cosmic rays spectrum. It leads also
   appearance  of limiting density dependent on energy which takes place only
  at very high energies and has an origin related to unitarity saturation.

\section*{Acknowledgments}
We are grateful to J.-R. Cudell, M. Islam, L. Jenkovszky, A. Krisch, U. Maor,
A. Martin, E.~Martynov, V. Petrov and D. Sivers for
the interesting comments and discussions.

\end{document}